\documentclass[3p,number,sort&compress]{elsarticle}
\usepackage{amsmath}
\usepackage{amsfonts}
\usepackage{amssymb}
\usepackage{graphicx}

\usepackage{indentfirst}

\begin{document}
\begin{frontmatter}


\title{Simulated annealing algorithm for finding periodic orbits of multi-electron atomic systems}

\author[CPT]{F. Mauger}
\author[CPT]{C. Chandre} 
\author[GA]{T. Uzer}
\address[CPT]{Centre de Physique Th\'eorique, CNRS -- Aix-Marseille Universit\'e, Campus de Luminy, case 907, F-13288 Marseille cedex 09, France} 
\address[GA]{School of Physics, Georgia Institute of Technology, Atlanta, GA 30332-0430, USA}
\date{\today}

\begin{abstract}
We adapt the simulated annealing algorithm to the search of periodic orbits for classical multi-electron atomic systems. This is done by minimizing the $n$-th return distance to the initial position on a Poincar\'e surface of section under an energy constraint. Here we give evidence of the feasibility of the method by applying it to the helium atom in the ground state for one to three spatial dimensions. We examine the structure of the dynamics and connect its organization to the periodic orbits we have found.
\end{abstract}

\end{frontmatter}


\section{Introduction}

Periodic orbits play a central role in the description and analysis of dynamical systems as they represent the skeleton of the dynamics~\cite{chaosbook}. It means that some important dynamical properties can be deduced from these orbits. By continuity of the flow a typical trajectory approaching a periodic orbit will mimic its dynamics. The time delay during which the trajectory is caught by the periodic orbit depends on the stability of the orbit: the more stable the periodic orbit, the longer the trajectory will stay in its neighborhood. As a result, the knowledge of the periodic structures and their stability properties enables one to predict the dynamical organization of the flow in their neighborhood. A fair amount of information is provided by the linear stability properties of these invariant structures. As an example, cycle expansions according to the length, stability or action of these orbits, are carried out to describe long time behavior such as averages of observables~\cite{Lan10,chaosbook}. For chaotic systems, the symbolic dynamics describes a hierarchy between periodic orbits, and the resulting expansion converges exponentially or superexponentially with the cycle length. 
For autonomous Hamiltonian systems, the eigenvalues of the Jacobian matrix from which the linear stability properties are determined come in quadruplet $(\lambda,1/\lambda,\lambda^*,1/\lambda^*)$. In addition, there are at least two marginal eigenvalues corresponding to the time translation invariance (along the periodic orbit) and the energy conservation. These eigenvalues allow the classification of periodic orbits according to their linear stability properties. For instance,
for Hamiltonian systems with two degrees of freedom, the periodic orbits can be sorted in three categories depending on their linear stability property: they are either elliptic (in general, linearly stable), hyperbolic (linearly unstable) or parabolic (linearly neutral). Elliptic periodic orbits are generally surrounded by an elliptic island inside which trajectories are trapped (and stay on invariant tori). In this region, the dynamics is mainly ruled by the central periodic orbit and the size of the island is determined by nonlinear stability properties. For hyperbolic periodic orbits, neighboring trajectories (linearly or locally) exponentially diverge in time. In general, these orbits are surrounded by a chaotic dynamics of stretching and folding since their stable and unstable manifolds intersect an infinite number of times to create a chaotic tangle.

Of course the influence of various kinds of periodic orbits strongly depends on the problem at hand. For atomic and molecular systems, the typical duration of a process is short compared to the other physical processes. This is particularly true for systems driven by short and intense laser pulses. Periodic orbits longer than this typical duration (or even of the same order) will not influence drastically the dynamics. On the contrary, periodic orbits much shorter than this duration will have a chance to trap the trajectory in its neighborhood and hence significantly affect the dynamical properties. 

Various algorithms have been developed for finding periodic orbits. Among them, some are commonly used in atomic systems with a few number of electrons~: Newton-Raphson algorithm or modified version of it such as the damped Newton-Raphson algorithm offers high speed convergence provided one has a sufficiently accurate initial guess for the periodic orbit. The precision of this initial guess is a thorny problem since the basin of attraction shrinks exponentially with the instability and the length of the orbit~\cite{davi99,Koh07}. A solution to overcome this difficulty is to consider a parallelized version of the algorithm through a multi-shooting strategy~\cite{chaosbook}. Independently of the chosen version, a crucial point in Newton-Raphson methods is the evaluation of the derivative of the trajectory which can be a delicate point from a numerical point of view. For a flow with dimension $N$, the tangent flow is given by the evolution of a $N\times N$ matrix which increases significantly the dimensionality of the problem at hand (even if in some cases, the structure of the problem may enable one to reduce the actual number of components). Some methods were set up to overcome this issue, and are generally based on relaxation algorithms~\cite{Biha89,Schm97,Diak98,davi99,davi01}. Overall, the goal is to set up a deterministic dissipative dynamical system which has the periodic orbit as its attractor (with hopefully a wide basin of attraction). Using the same philosophy, variational methods have been designed to determine periodic orbits. For instance, a variational principle combined with a Newton descent~\cite{Lan04} consists in setting up a (dissipative) fictitious-time dynamics in a space of loops such that it drives a loop to a true periodic orbit of the considered dynamical system. Similarly some other algorithms are designed based on minimizing a function (or a functional). For instance one can consider a Newton-Gauss method by looking at minimization of the distance between the starting and final points of an orbit on a particular surface intersecting the periodic orbit: Global vanishing minima correspond to periodic orbits. 

Independently of the chosen algorithm, the chance to see the algorithm converging strongly depends on the basin of attraction of the periodic orbit of the deterministic dynamical system. Often, there is a trade-off between the size of the basin of attraction with the speed of convergence of the algorithm. This is in particular the case for the Newton-Raphson and its damped version. In this article we propose a method which extends this basin of attraction by use of systematized trial and error converging procedure.  In order to do that, we combine these deterministic algorithms with a Simulated Annealing (SA) algorithm to approximate the periodic orbits. In other words, the goal of this stochastic method is to enable the determination of initial guesses with sufficient accuracy to lay them into the basin of attraction of a fast converging algorithm like the Newton-Raphson algorithm. As a consequence of the underlying stochastic nature of the algorithm, it enables one to determine several different periodic orbits for the considered dynamical system by launching the algorithm several times. One of the advantages of the SA algorithm is that it does require neither the computation of the tangent flow nor a high accuracy in the integration of the trajectory. We show below that it makes this method a tool of choice for systematic search for periodic orbits in phase space.

To illustrate the feasibility of the SA algorithm for finding periodic orbits, we consider the classical helium atom in its ground state.
The dynamics is modeled by the following Hamiltonian with soft Coulomb potentials~\cite{Java88}:
\begin{equation} \label{eq:Hamiltonian}
   {\mathcal H} \left( {\bf x}_1, {\bf x}_2, {\bf p}_{1}, {\bf p}_{2} \right) =
      \frac{ \left| {\bf p}_{1} \right|^{2}}{2} + \frac{ \left| {\bf p}_{2} \right|^{2}}{2} 
      - \frac{2}{\sqrt{ \left| {\bf x}_1 \right|^{2} + a^{2}}} - \frac{2}{\sqrt{ \left| {\bf x}_2 \right|^{2} + a^{2}}}
      + \frac{1}{\sqrt{ \left| {\bf x}_1 - {\bf x}_2 \right|^{2} + b^{2}}}.
\end{equation}
where ${\bf x}_i$ is the position of the $i$-th electron (the nucleus being set at the origin), and ${\bf p}_{i}$ is its canonically conjugate momentum. The norm $\left| \cdot \right|$ is the Euclidian norm. Depending on the dimensionality of the model, the vectors ${\bf x}_i$ and ${\bf p}_i$ can be considered in $\mathbb{R}$, $\mathbb{R}^{2}$ or $\mathbb{R}^{3}$. Tuning the softening parameters $a$ and $b$ enables to reproduce various atomic configurations. The parameter $a$ is chosen as to reproduce the ground state energy (defined as the sum of the first and second ionization potentials) while preventing any self ionization of the atom. The parameter $b$ is chosen as to allow significant energy exchange between the electrons when they are close to each other. For helium, one usually considers $a = b = 1$ and the ground state energy is $E_{g}=-2.24$ a.u.~\cite{Haan94}.

The dynamical organization corresponding to Hamiltonian~(\ref{eq:Hamiltonian}) is crucial for understanding the (multiple) ionization processes of these atoms driven by an intense and short laser pulse. In Ref.~\cite{Maug09_1} it was argued from a one dimensional model that the dynamics of Hamiltonian~(\ref{eq:Hamiltonian}) is mainly organized by a reduced set of periodic orbits which naturally places one electron close to the nucleus and one further away. This distinction between the two electrons is crucial when understanding the action of an intense linearly polarized laser field when it is applied to the system~: The electron far away from the nucleus is picked up by the field and hence quickly ionizes while the other electron remains bound to the nucleus. This distinction between the two electrons is at the origin of the classical interpretation of the recollision or the "`three-step"' model which is the keystone of strong-field physics~\cite{Cork93,Scha93,Beck08}.
The question we are addressing here is whether this emerging picture of an inner (close to the nucleus) and an outer electron (further away) still exists in two or three dimensions, or if it is just an artifact of the one dimensional model which is very peculiar given the presence of the nucleus. We show below that this dynamical picture still persists and and it does so because of the organization of short periodic orbits in phase space. 

In Sec.~\ref{sec:1}, we describe the algorithm we use to determine periodic orbits. In Sec.~\ref{sec:2}, we analyze the lay-out of these periodic orbits in phase space and their implication for the dynamical organization of the helium atom in the ground state. 


\section{Simulated annealing algorithm to determine periodic orbits} 
\label{sec:1}

The Simulated Annealing (SA) algorithm~\cite{Kirk83} is a metaheuristic algorithm designed for optimization under constraints which generalizes the ideas developed from the Metropolis algorithm~\cite{Metr53}. The main idea of the algorithm is to automate a trial and error search for a better solution within a controlled neighborhood whose diameter is successively reduced. 

Given a dynamical system $\dot{\bf X} = {\bf F} \left( {\bf X} \right)$, we first consider a Poincar\'e section $\Sigma$ and the associated Poincar\'e map $\Phi : \Sigma \mapsto \Sigma$. If a trajectory cuts $\Sigma$ in a finite set of points $\left( {\bf X}_{1}, {\bf X}_{2}, \ldots, {\bf X}_{n} \right)$ such that $\Phi \left( {\bf X}_{n} \right) = {\bf X}_{1}$ and $\Phi \left( {\bf X}_{i} \right) = {\bf X}_{i+1}$, $1 \leq i < n$ it corresponds to a periodic orbit with period $n$ on the section (in other words $\Phi^{n}\left({\bf X}_{1}\right)={\bf X}_{1}$). The periodic orbit search is obtained by minimizing the distance between the starting point and the $n$-th return on $\Sigma$ using the SA algorithm. From a random initial guess ${\bf X}^{(0)}\in \Sigma$ we build up a sequence $\left\{{\bf X}^{(k)}\right\}_{k\in\mathbb{N}}\in \Sigma^{\mathbb N}$ which aims at minimizing the Euclidean distance between ${\bf X}^{(k)}$ and $\Phi^{n}\left({\bf X}^{(k)}\right)$. If the SA algorithm converges, the limit ${\bf X}^{\left(\infty\right)}$ of the sequence exists and fulfills the condition $\Phi^{n} \left({\bf X}^{\left(\infty\right)}\right) = {\bf X}^{\left(\infty\right)}$, therefore corresponding to a periodic orbit (with period $n$ on $\Sigma$).
For the problem at hand, finding periodic orbits of Hamiltonian~(\ref{eq:Hamiltonian}) in the ground state, an additional constraint to take care of during the SA algorithm is the energy of the system (which is constant for autonomous Hamiltonian systems). There are several ways of doing it~: The first one is to ensure that the points ${\bf X}^{(k)}$ remain on the energy surface after each step of the SA algorithm by an additional procedure which adjusts one coordinate. The second way is a specific prescription on the function to minimize which only ensures that the limit ${\bf X}^{(\infty)}$. 

For a given function $f: \Gamma \rightarrow \mathbb{R}$, where $\Gamma$ is a subset of $\Sigma$, called cost function, the goal is to approximate the global minimum of $f$ over $\Gamma$.
First a random initial condition~${\bf X}^{(0)} \in \Gamma$ is chosen to initiate the SA algorithm and we define two parameters $T_{0}$, the initial temperature, and $\kappa\in\mathbb{R}^{+}$ which serves as error tolerance.

The SA algorithm proceeds in successive stages of annealing and cooling steps.
\begin{itemize}
   \item During the {\em annealing} process, a random initial guess ${\bf X}^{(1)}$ is chosen in the neighborhood of ${\bf X}^{(0)}$ in $\Gamma$ (the size of the neighborhood is controlled by $T_{0}$; for simplicity one usually chooses a ball with diameter $T_{0}$). It corresponds to a melting of the system with a controlled temperature $T_{0}$. If $f \left( {\bf X}^{(1)} \right) < f \left( {\bf X}^{(0)} \right)$ then ${\bf X}^{(1)}$ replaces ${\bf X}^{(0)}$ in the following step. If not, ${\bf X}^{(1)}$ replaces ${\bf X}^{(0)}$ with a probability $\exp ( -(f( {\bf X}^{(1)}) - f( {\bf X}^{(0)}))/\kappa T_{0})$. Otherwise, the same ${\bf X}^{(0)}$ is kept in the next step. The process of melting for a fixed temperature is iterated until the system reaches a steady state.
   \item The {\em cooling} consists in decreasing the temperature to a new temperature $T_{1} < T_{0}$, realized at the end of the annealing process. Then a new iteration of annealing with the updated temperature $T_{1}$, is performed. 
\end{itemize}
The successions of annealing and cooling steps are iterated until the system reaches a global steady state which is the resulting approximation of the global minimum. Since the size of the neighborhood where the perturbation is performed is controlled by the temperature, it has to shrink with iterations of cooling. At the limit for zero temperature, the neighborhood of control is reduced to a single point.

The constant $\kappa$ is used to avoid the algorithm to be trapped in local minima. It has to be chosen carefully as two critical situations may arise:
\begin{itemize}
   \item If $\kappa$ is too large, the algorithm will not converge since too many ``bad'' perturbations are allowed. 
   \item If $\kappa$ is too small, one runs the risk to be trapped in local minima. 
\end{itemize}
The cooling law also has to be chosen carefully to allow the algorithm to sufficiently explore the region around the initial guess. If the cooling is too fast, the system is frozen in its current position and the algorithm may not converge. If the cooling is too slow, the algorithm spends a lot of time in useless explorations which slows down the process. Usually, an exponential decrease is considered for the cooling by multiplying the current temperature with a constant, i.e.\ $T_{n+1} = \alpha T_{n}$ where $0 < \alpha < 1$. 

Recall that our goal is to find periodic orbits for Hamiltonian~(\ref{eq:Hamiltonian}) with the constraint to be on the ground state, and to do that, we are looking for $n$ periodic points on a Poincar\'e surface. For instance, we choose the Poincar\'e surface $\Sigma$ of equation $x_{1} = 0$ with $\dot{x}_{1} > 0$, where $x_{1}$ is the first component of ${\bf x}$ in the canonical basis (other surfaces which intersect the ground state energy surface might also be suitable). For convenience, we denote ${\bf X} = \left( {\bf x}_1, {\bf x}_2, {\bf p}_{1}, {\bf p}_{2} \right)$ the vector position in phase space. Our problem is to find a periodic point ${\bf X}^{\left(\infty\right)}$ on the Poincar\'e (and the ground state energy) surface such that ${\bf X}^{\left(\infty\right)} = \Phi^{n}\left({\bf X}^{\left(\infty\right)}\right)$. We define the cost function as the $n$-th return distance function
\begin{equation} \label{eq:preturndistfunc}
   f : {\bf X} \mapsto \left| {\bf X} - \Phi^{n} \left( {\bf X} \right) \right|.
\end{equation}
such that periodic orbits with period $n$ correspond to positions where the global minimum of the function $f$ is reached and thus may be investigated using the SA algorithm. This function is defined on $\Gamma$ which is the intersection of the ground state energy surface $\left\{{\bf X} \ {\rm s.t.} \ {\mathcal H} \left({\bf X}\right) = E_{g} \right\}$ with the Poincar\'e section $\Sigma$.


Schematically, the algorithm is written as
\begin{verbatim}
% Initialization
    Xold      =   InitCond()                      % Random initial condition generation
    DistError =   ReturnDist(Xold)                % n return distance from initial condition
    CoolIter  =   0                               % Cooling iteration counter
    
% Simulated annealing algorithm
    while (CoolIter < #CoolIter) and (DistError < #DistError)
        MeltIter   =   0                          % Melting iteration counter
        
        while (MeltIter < #MeltIter) and (DistError < #DistError)
            Xnew   =   Xold + Tprt*2*(rand()-.5)  % (*) random perturbation of current guess
            Dist   =   ReturnDist(Xnew)           % n return distance for perturbed condition
            
            if Dist < DistError                   % Improved guess
                Xold      =   Xnew
                DistError =   Dist
                
            % Allowed not improving perturbation
            else if (exp(-(Dist-DistErr)/(#ErrTol*Tprt)) < rand())
                Xold      =   Xnew
                DistError =   Dist
                
            end if
            
            MeltIter = MeltIter + 1
            
        end while
        
        Tprt     =   Tprt*#TprtFact
        CoolIter =   CoolIter + 1
        
    end while
\end{verbatim}
where user-defined constants are labeled by \# and \% starts comments. The function \emph{ReturnDist} plays the role of the function $f$ as defined by Eq.~(\ref{eq:preturndistfunc}). Finally, the function \emph{rand} is a random generator uniformly distributed over $[0,1]$. The melting process $(*)$ is draffed in its simplest way above and it has to be followed by a projection on $\Gamma$, which has to be chosen accordingly to the problem at hand. 
In our case, after the melting $(*)$, the new  initial condition is neither on $\Sigma$ nor on the ground state energy. The projection on $\Sigma$ is simply done by setting $x_{1} = 0$. The projection on the ground state is done by changing $p_{x,1}$ accordingly to the energy constraint and we select the positive solution for $p_{x,1}$ in order to be sure to remain on the Poincar\'e section. We also define a threshold distance under which the algorithm is considered as converged.

The projection of the perturbed position on the ground state energy surface ensures that the series of guesses stay on the energy surface at each iteration. However, one can also see the ground state energy as an additional constraint to minimize together with the $n$-return distance. In this case, one adds the energy error to the cost function
\begin{equation}
\label{eq:const}
   f:{\bf X} \mapsto \left|{\bf X} - \Phi^{n}\left({\bf X}\right)\right| + \gamma \left| {\mathcal H}\left({\bf X}\right) - E_{g}\right|,
\end{equation}
where $\gamma > 0$ is a constant. Then, after each perturbation, we only project the new guess on the Poincar\'e surface $\Sigma$. We have successfully implemented both methods for $n\in\left\{1,2,3\right\}$. All numerical results displayed in this paper and the values given for the parameters correspond to the projection (on both the Poincar\'e and ground state energy surfaces) method.

Many improvements of the algorithm can be implemented depending on the problem at hand. For instance, there is no requirement for choosing the initial condition randomly. If one has an approximate guess for the location of a periodic orbit, or any subset of phase space where it is included (e.g., given the symmetries of the problem), it is more suitable to start the SA algorithm with this particular point. In such a case, it can be interesting to set the initial temperature cooler than for a random initial guess. 

For improving the convergence, it is more efficient to combine the SA algorithm with a second method based on a Newton-Raphson method for instance. Here we use the function \texttt{fsolve} of Matlab, which is a Newton-Gauss minimizing scheme.

\section{Application to two electron atoms with soft Coulomb potentials}
\label{sec:2}

For two spatial dimensions the complexity of the system is increased compared to one dimensional models since the number of degrees of freedom is increased from 2 to 4. In addition, for two spatial dimensions, the system has a continuous symmetry which corresponds to a global invariance by rotation of the atom in phase space. The associated conserved quantity is the total angular momentum ${\bf x}_1 \times {\bf p}_{1} + {\bf x}_2 \times {\bf p}_{2}$, where $\times$ is the cross product in $\mathbb{R}^{2}$: ${\bf x} \times {\bf p} = x p_y - y p_{x}$. It means that any periodic orbit generates an infinite family of periodic orbits deduced from this single one by a rotation of the atom in phase space. In addition, the system also has some discrete symmetries (the exchange of the role of the two electrons for instance). To simplify the analysis, in what follows, we identify as a single one all periodic orbits which can be deduced by one of those symmetries from another periodic orbit. We notice that these periodic orbits have the same period and the same linear stability properties. In the numerical implementation, we consider that a periodic orbit has converged when the distance in phase space between the starting and final points on the Poincar\'e section is smaller than $10^{-10}$. We consider a periodic orbit to be a new one when the difference in period or in the spectrum of the monodromy matrix is larger than $10^{-2}$ from any already known periodic orbit in the list.

For actual use of the SA algorithm we set the initial temperature around 5. A tolerance parameter $\kappa = 0^{+}$ (a perturbation is accepted if an only if it reduces the cost function) is enough for our problem when we perform the projection on the ground state energy surface. It should be noted that if we add the energy error to the cost function [see Eq.~(\ref{eq:const})], $\kappa \approx 5\times 10^{-2}$ is a better choice. For the cooling we set the temperature ratio to $\alpha = 0.75$. Finally, we iterate the melting step a fixed number of times (3000 melting steps during the annealing process for each temperature). We stop the cooling when the temperature has reached a threshold value (here the threshold temperature is $10^{-5}$). We also define a threshold $d_{crit}$ for which the algorithm is stopped as soon as the current guess ${\bf X}$ satisfies $f \left( {\bf X} \right) < d_{crit}$, and here $d_{crit} = 10^{-3}$.

For actual use of the SA, there are two main strategies when choosing the temperature ratio and the number of melting steps. One strategy is to choose a fast cooling law (i.e. $\alpha$ small) and a large number of melting steps for each annealing. Another strategy is to consider a slow cooling rate (i.e. $\alpha \approx 0.95$) and a much reduced number of melting per annealing iteration. For the problem at hand, we have used the first strategy when considering a projection on the ground state energy surface, and we have used the second strategy when adding the energy error to the cost function.

The total angular momentum is a conserved quantity of the dynamics and to model a configuration closer to the quantum ground state, one may wish to impose it to be zero. That becomes a new constraint for our system and it decreases the actual dimensionality of the problem at hand, but still it is worth looking for periodic orbits which may organize this particular subclass of the dynamics. To do so, there are two straightforward ways to adapt the algorithm presented here. The first one consists in adding the total angular momentum to the cost function (in the same way as for the energy constraint)
$$
   f: {\bf X} \mapsto \left| {\bf X} - \Phi^{n} \left( {\bf X} \right) \right| + \gamma \left| {\bf x}_1 \times {\bf p}_{1} + {\bf x}_2 \times {\bf p}_{2} \right|,
$$
where $\gamma$ is constant. The total angular momentum is then considered as an additional constraint and the SA will try to minimize it together with the $n$-th return distance function. Another solution is to project, after each melting iteration, the guess on the intersection of the ground state energy and zero total angular momentum surfaces. We have successfully adapted our algorithm to find one dimensional, three dimensional and zero total angular momentum two dimensional (considering the projection alternative) periodic orbits.

Since we are interested in finding as many periodic orbits as possible up to a finite period due to the finite duration of laser pulses in experiments, we launch a large number of times the algorithm, to explore the whole accessible phase space. We do it for periodic orbits of Hamiltonian~(\ref{eq:Hamiltonian}) which have up to three intersections on the Poincar\'e section.  
Because of the rotational invariance of the system, the monodromy matrix associated with a periodic orbit has an additional pair of eigenvalues 1 (in addition to the pair associated with time-translation along the orbit). As a result, for two spatial dimensions, four eigenvalues are equal to one in the monodromy matrix and the four others determine the linear stability properties. 
Consequently the system can exhibit nine kinds of linear stability, (elliptic, parabolic or hyperbolic)-(elliptic, parabolic or hyperbolic). However, for numerical computations, is it difficult to distinguish precisely the parabolic features, e.g., between a parabolic and an elliptic or a weakly hyperbolic periodic orbit. When checking the linear stability of the periodic orbits we have found, we arbitrarily group together parabolic and elliptic features (it is worth noting that we have found no fully parabolic periodic orbit).

\begin{figure}
	\centering
		\includegraphics[width = .5\linewidth]{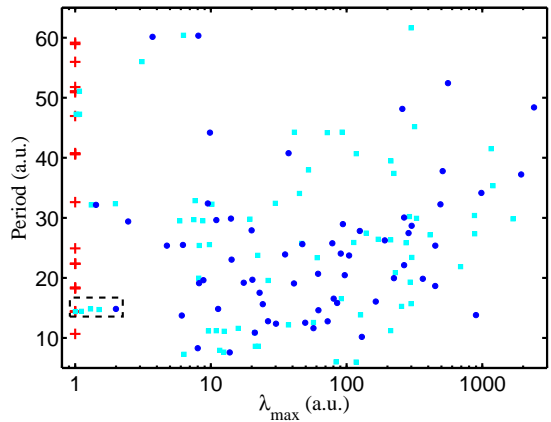}
	\caption{\label{fig:StabPer}
	 Period $T$ versus the norm of the largest eigenvalue $\lambda_{\rm max}$ of the monodromy matrix for the detected periodic for Hamiltonian~(\ref{eq:Hamiltonian}) with two spatial dimensions. Elliptic-elliptic and parabolic-elliptic periodic orbits are labeled with red crosses, elliptic-hyperbolic and parabolic-hyperbolic ones with cyan squares and hyperbolic-hyperbolic ones with blue circles. Hyperbolic periodic orbits into the dashed black rectangle are depicted in Fig.~\ref{fig:PO}.}
\end{figure}

For the case of two spatial dimensions for Hamiltonian~(\ref{eq:Hamiltonian}), we have found a total of 155 periodic orbits with period smaller than 65 a.u.\ whose repartition is the following one~: 43 periodic orbits with period one, 74 periodic orbits with period two (which cannot be reduced to period one, that is to say, prime periodic orbits) and 38 periodic orbits with period three. Regarding their stability properties, 22 are elliptic-elliptic or parabolic-elliptic, 71 are elliptic-hyperbolic or parabolic-hyperbolic and 62 are hyperbolic-hyperbolic. In Fig.~\ref{fig:StabPer}, we display the period $T$ as a function of the modulus of the largest eigenvalue $\lambda_{\rm max}$ of the monodromy matrix for each periodic orbit determined by the algorithm. By no means are these periodic orbits the only ones in phase space. However, these are the ones obtained by launching a large number of times the SA algorithm (with the restrictions mentioned above).

\begin{figure}[htb]
	\centering
		\includegraphics[width = .32\linewidth]{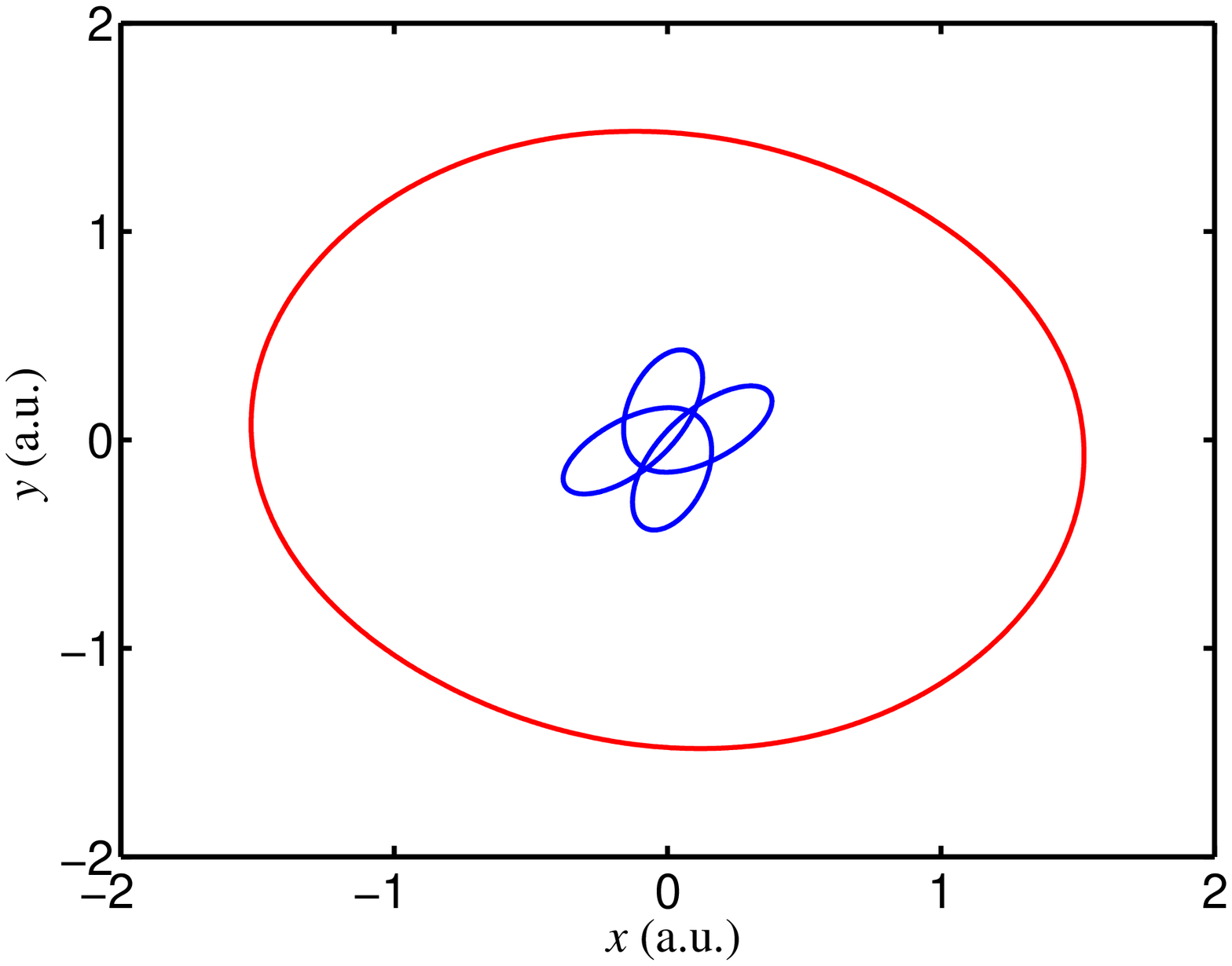}
		\includegraphics[width = .32\linewidth]{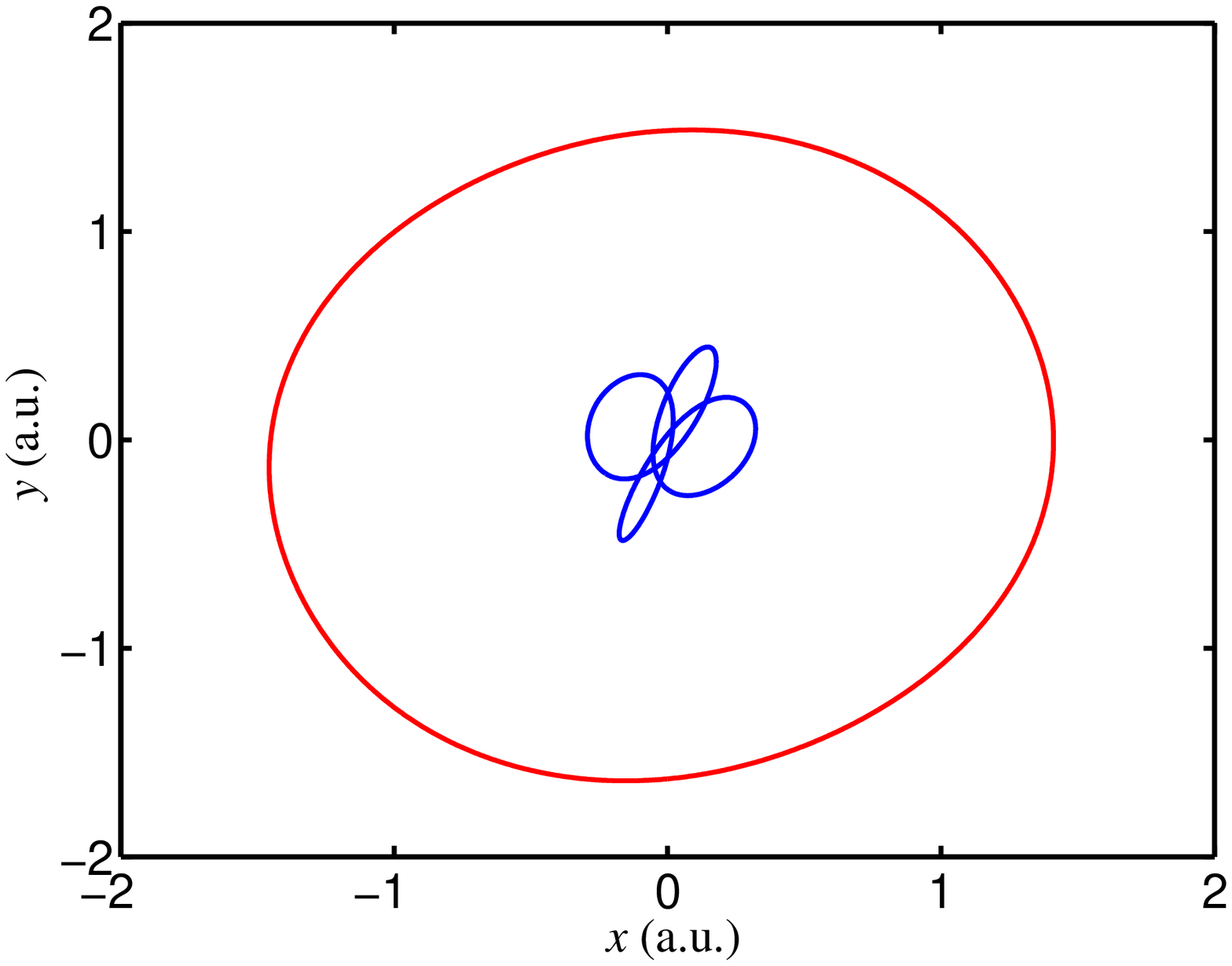}
		\includegraphics[width = .32\linewidth]{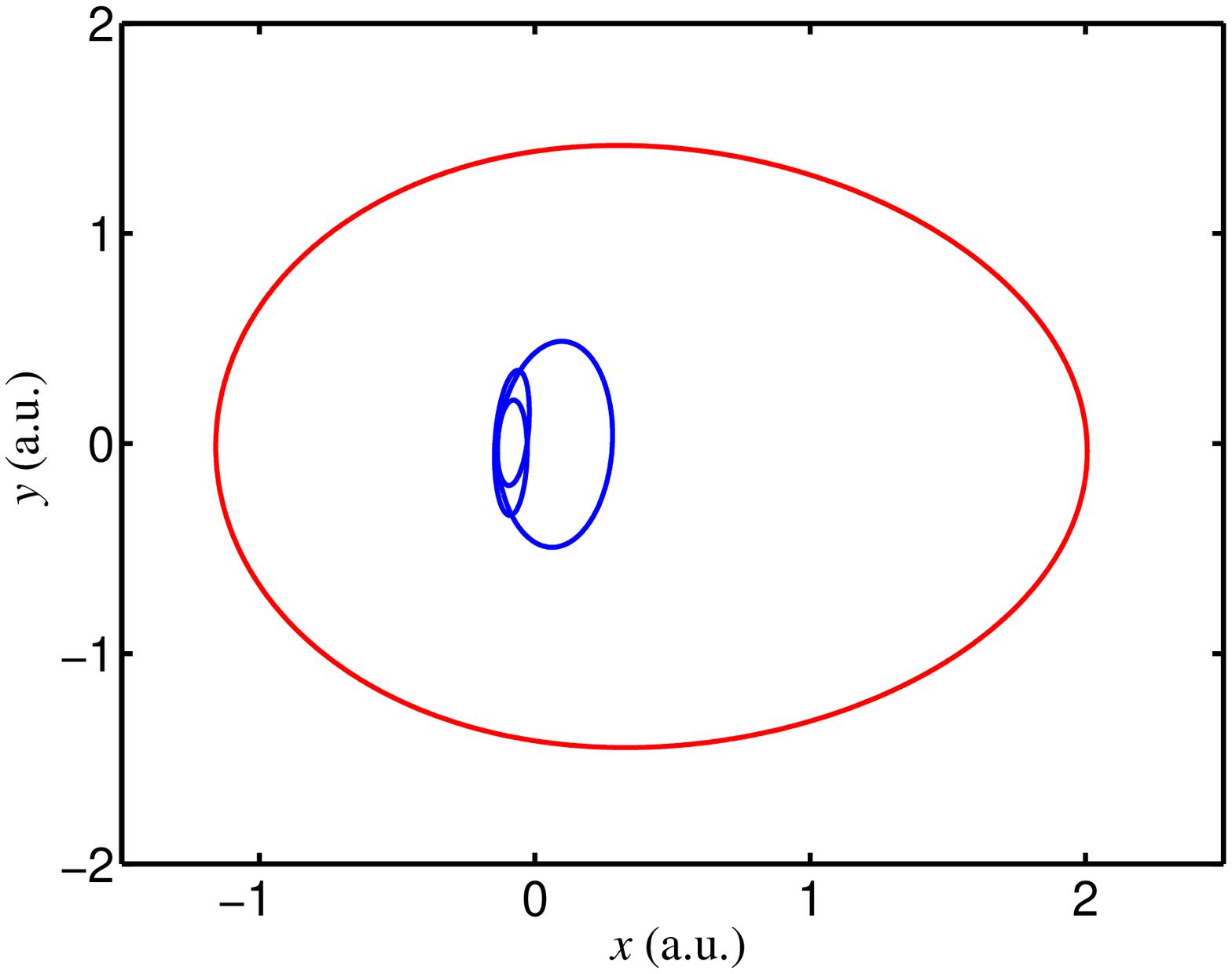}
		\includegraphics[width = .32\linewidth]{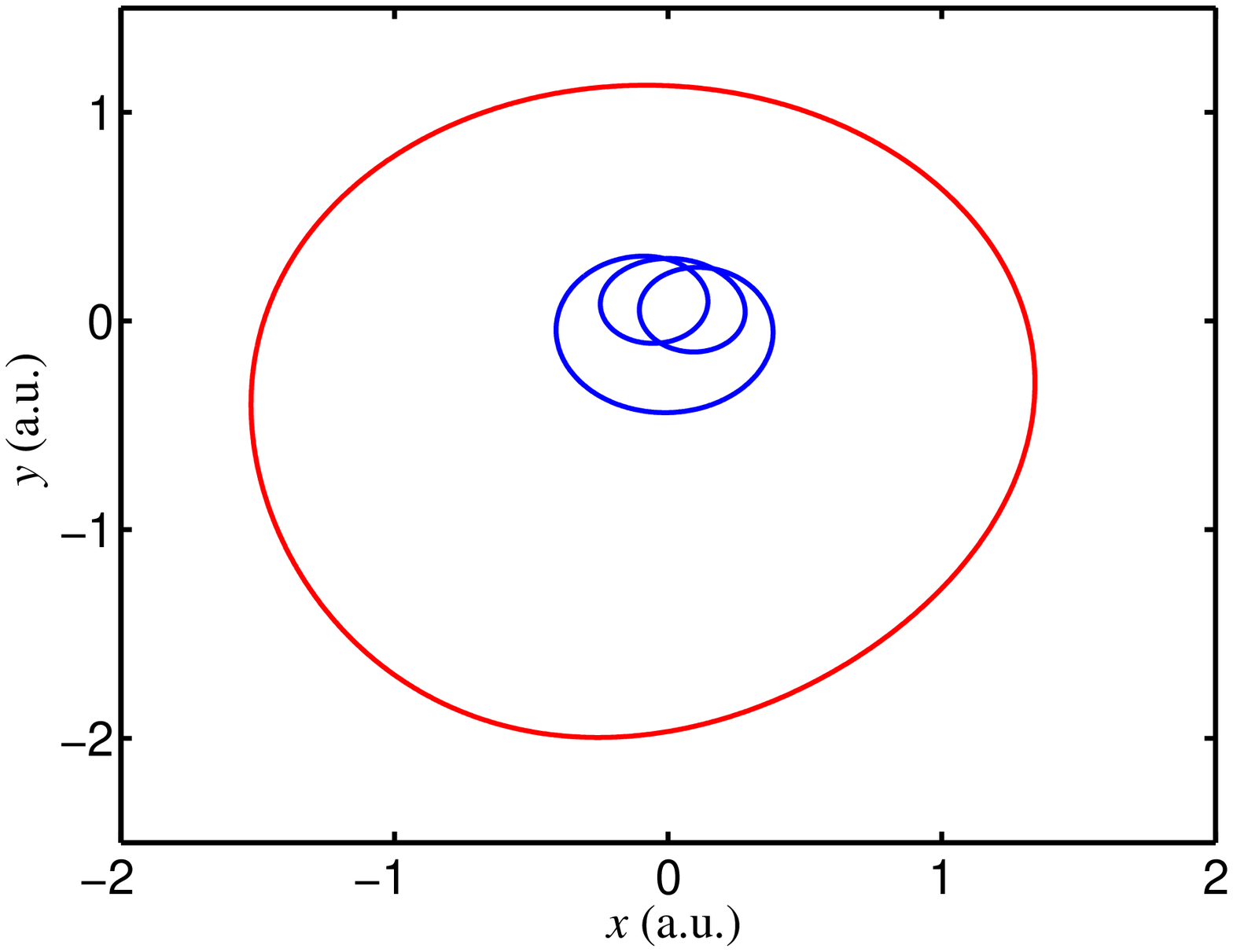}
		\includegraphics[width = .32\linewidth]{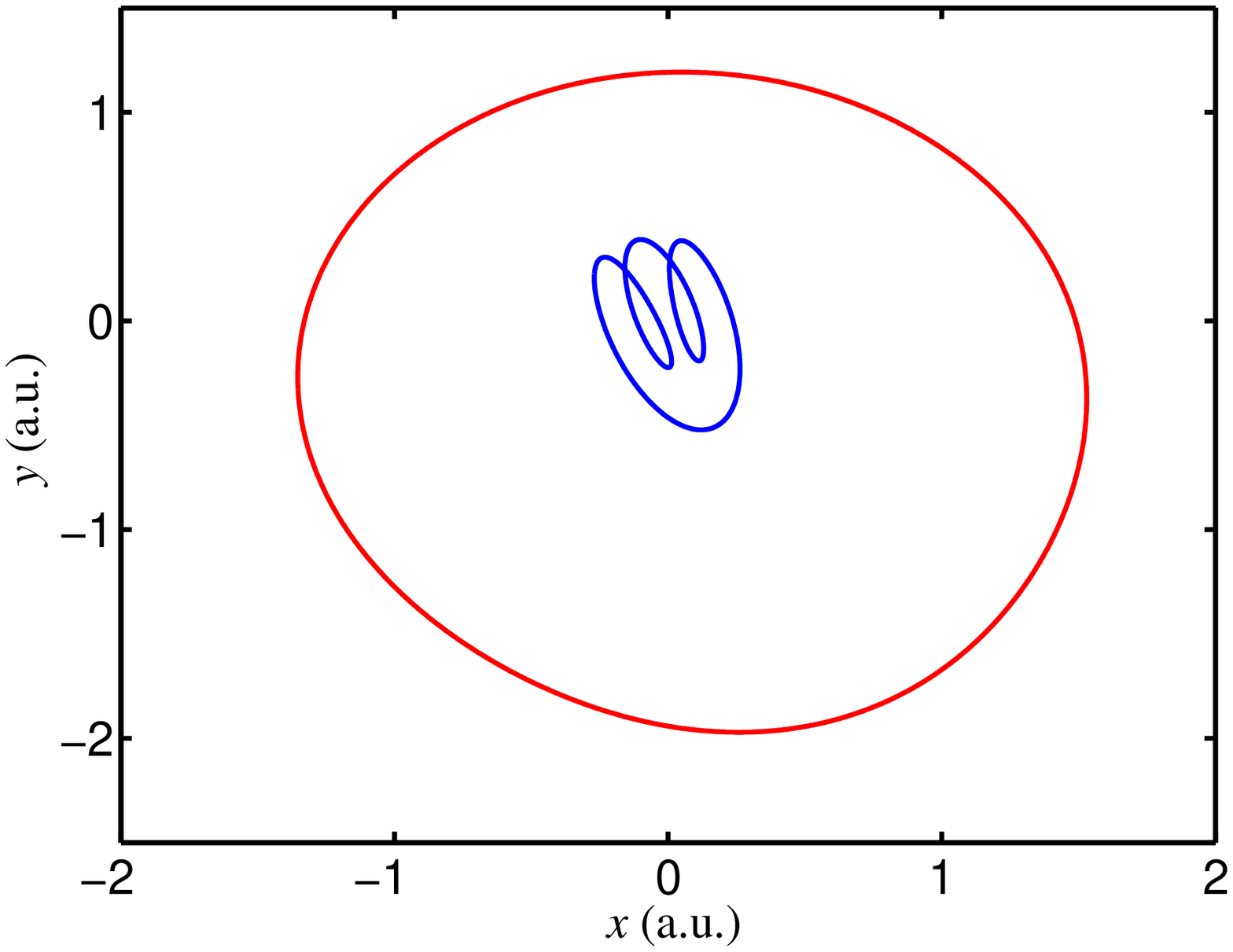}
	\caption{\label{fig:PO}
	Projection in the two dimensional configuration plane $(x,y)$ of the five hyperbolic periodic orbits inside the dashed black rectangle in Fig.~\ref{fig:StabPer}. Periodic orbits are sorted in increasing norm of the largest eigenvalue of the monodromy matrix, from left to right and up to down. Each electron is labeled with a different color (blue or red).}
\end{figure}

\begin{figure}[htb]
	\centering
		\includegraphics[width = .49\linewidth]{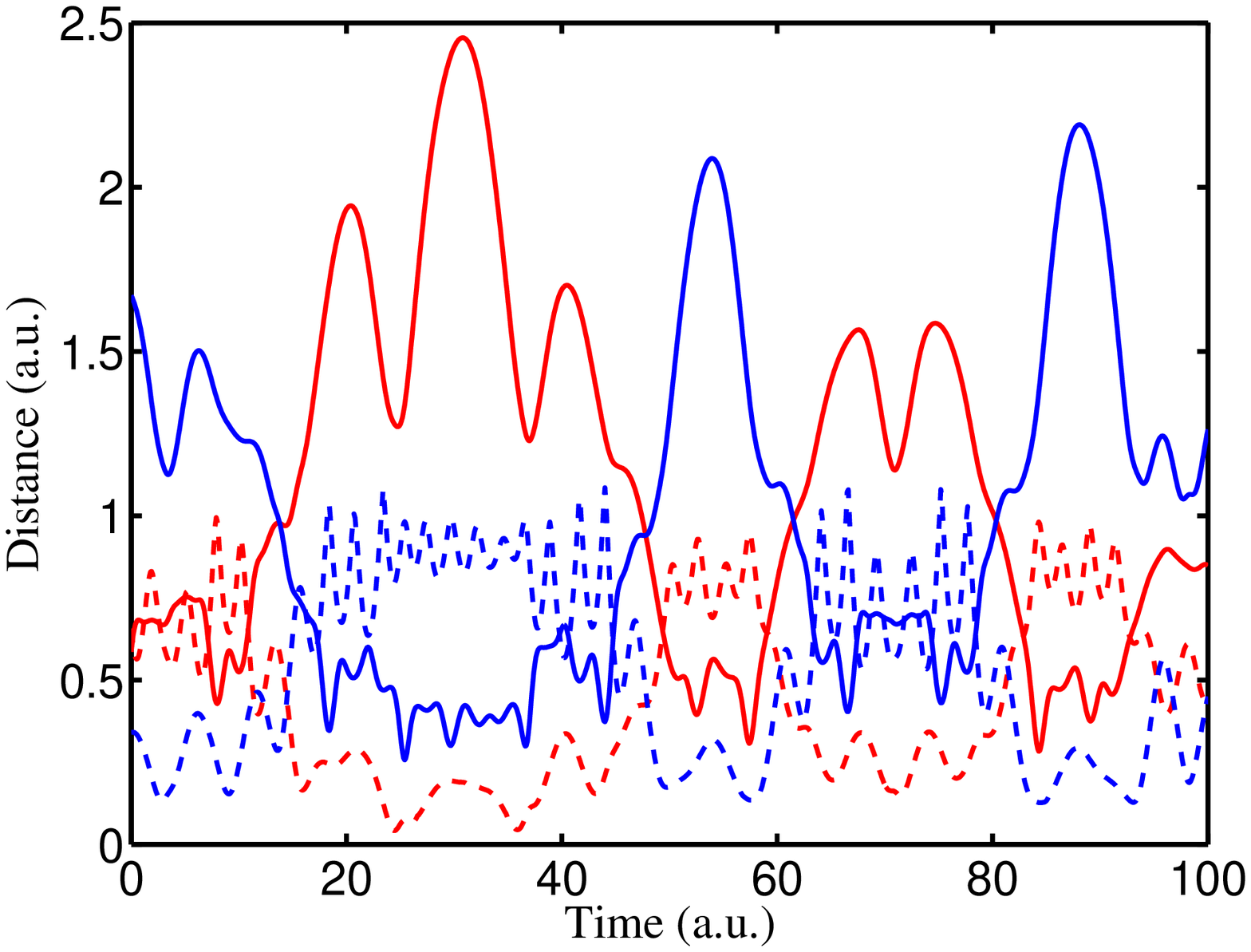}
		\includegraphics[width = .49\linewidth]{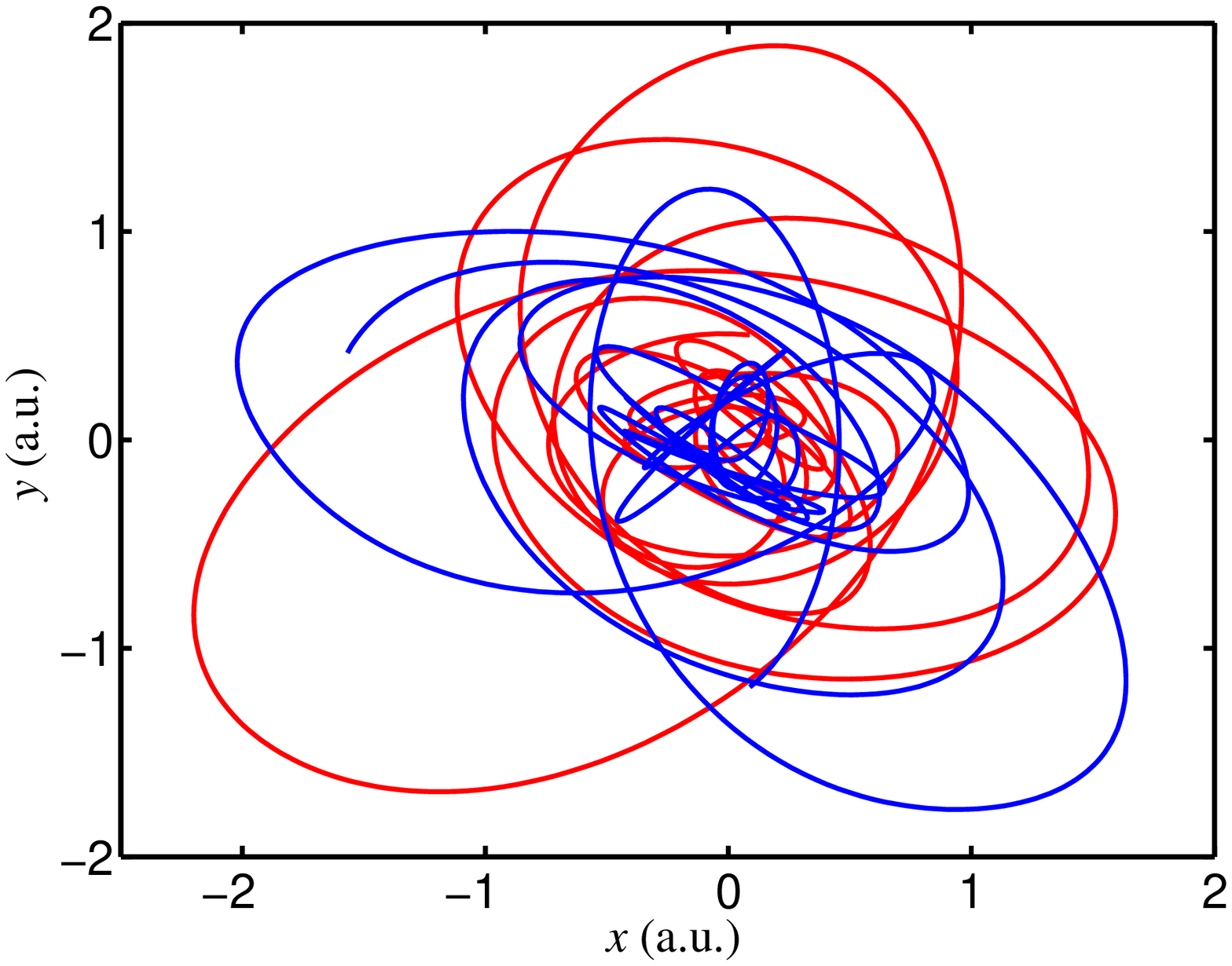}
	\caption{\label{fig:InnerOuter}
	Typical trajectory for Hamiltonian~(\ref{eq:Hamiltonian}) with two spatial dimensions for $a=b=1$, for a typical trajectory over the admissible phase space. On both panels, each electron is labeled by a constant color (red or blue). Left panel: Distance, in phase space, of each electron from the nucleus (continuous curves) and from the edge of the accessible phase space (dashed curves). Right panel: Projection of this trajectory in configuration space $(x,y)$.}
\end{figure}

There are five periodic orbits which are weakly hyperbolic and with relatively low period, and which are likely to organize the dynamics of Hamiltonian~(\ref{eq:Hamiltonian}) (see dashed black rectangle in Fig.~\ref{fig:StabPer}). This is so because all of those periodic orbits have a small period (around 15 a.u.) which is consistent with the short duration of laser pulses (typically of a few hundreds of a.u.\ which correspond to few tens of fs) and are weakly hyperbolic. These five hyperbolic periodic orbits are well separated, in the $(\lambda_{\rm max},T)$ diagram, from the other periodic orbits which are significantly longer and more hyperbolic. We display these five periodic orbits in the configuration space $(x,y)$ in Fig.~\ref{fig:PO}. We notice that all of them are composed of one electron close to the nucleus and another further away. A closer inspection at Fig.~\ref{fig:StabPer} also shows a group of four weakly hyperbolic periodic orbits with a larger period (around 30 a.u.). Looking at their shape in phase space (not shown here) they are also composed of one electron close to the nucleus and the other one further away. It means that for a typical trajectory of Hamiltonian~(\ref{eq:Hamiltonian}) under the influence of these periodic orbits it is possible to identify, at any time, an inner (close to the nucleus) and an outer (further away) electron with possible exchanges of the roles of the two electrons as the trajectory visits different areas of influence.
The observation that all small period and weakly hyperbolic periodic orbits are clearly composed of an inner and an outer electrons, supports what we have shown for the model with one spatial dimension where we identified four dominant periodic orbits in the organization of phase space (which actually reduce to only one orbit by symmetry)~\cite{Maug09_1,Maug09_2}. However, for models with two spatial dimensions, the variety of periodic orbits and the invariance by rotation make it difficult to identify a reduced number of dominant orbits in the skeleton of the dynamics. In order to show that a typical trajectory has an inner and an outer electron, we consider two distances in phase space as functions of time~: one from the nucleus and the other one from the boundary of the admissible phase shape for a typical trajectory of Hamiltonian~(\ref{eq:Hamiltonian}) as shown in Fig.~\ref{fig:InnerOuter}. At any time an inner and an outer electron are identified with quick and frequent exchanges of their roles~: The inner electron is close to the nucleus and thus have a short distance from it while the outer one, further away will be closer to the edge of the accessible phase space. It is worth noticing that due to the exchanges of the roles of the two electrons, the invariance by rotation and the dimensionality of the problem, a projection on the configuration plane does not clearly show any significant organization of the dynamics, contrary to the case with one spatial dimension which clearly shows the specific role played by mainly one periodic orbit (and its symmetric orbits). 

\begin{figure}[htb]
	\centering
		\includegraphics[width = .49\linewidth]{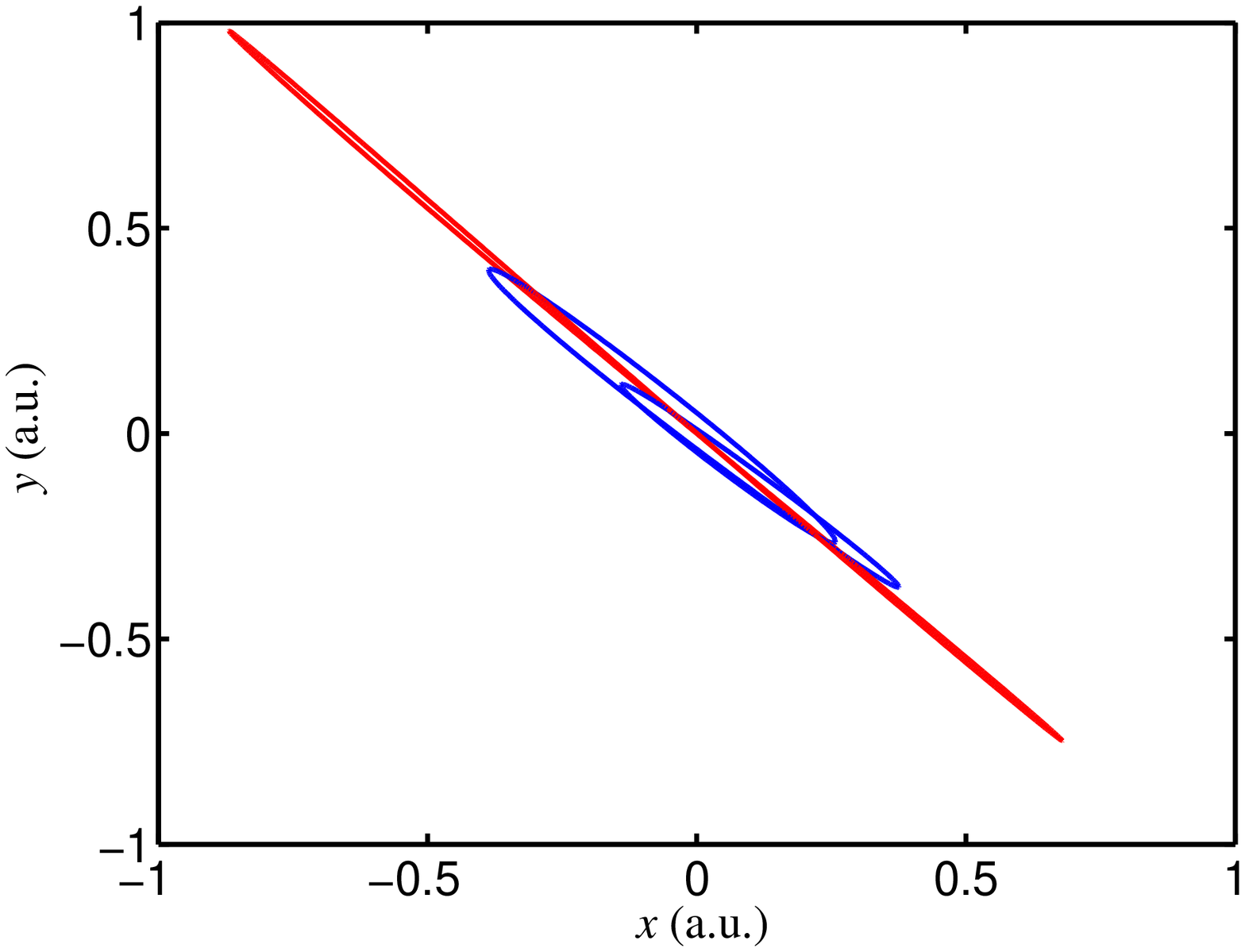}
		\includegraphics[width = .49\linewidth]{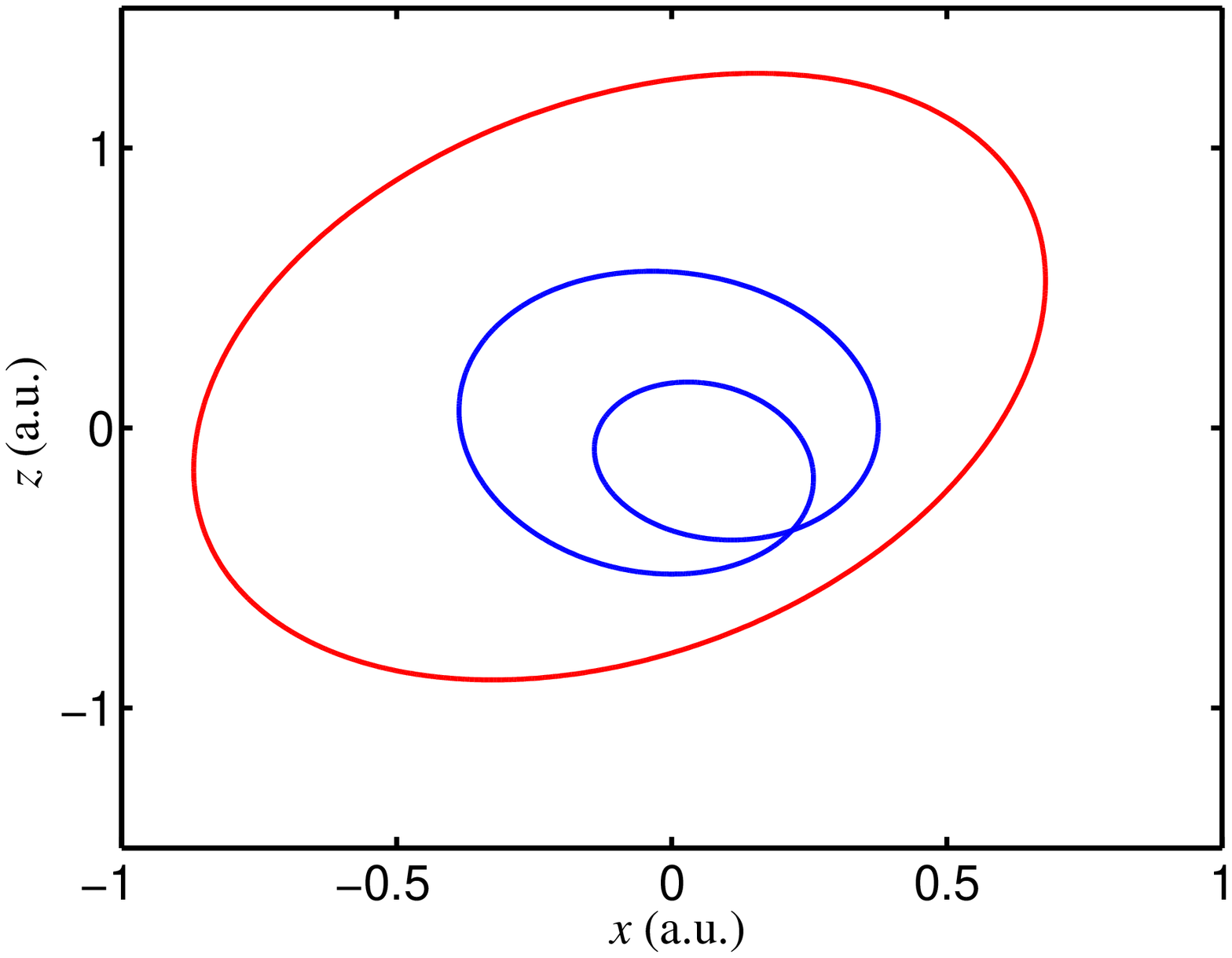}
	\caption{\label{fig:PO_3D}
	Projections on the $(x,y)$ plane (left panel) and $(x,z)$ plane (right panel) of a periodic orbit of Hamiltonian~(\ref{eq:Hamiltonian}) with three spatial dimensions.}
\end{figure}

The same study can be carried out in three spatial dimensions. The numerical results suggest that in most of the periodic orbits, the two electrons are co-planar or nearly co-planar (but not all). One of them is depicted in Fig.~\ref{fig:PO_3D}. These periodic orbits display a similar organization with one electron close to the nucleus and one further away.  
As a consequence, from a systematic inspection of the periodic orbits in phase space, we are able to draw a picture of the dynamics consisting of one electron close to the nucleus and another one further away (with fast exchanges between the two), confirming a result obtained with a reduced system with only one spatial dimension.  

\section*{Acknowledgments}
C.C. and F.M. acknowledge financial support from the PICS program of the CNRS. This work is partially funded by NSF.



\begin{thebibliography}{10}
\expandafter\ifx\csname url\endcsname\relax
  \def\url#1{\texttt{#1}}\fi
\expandafter\ifx\csname urlprefix\endcsname\relax\def\urlprefix{URL }\fi
\expandafter\ifx\csname href\endcsname\relax
  \def\href#1#2{#2} \def\path#1{#1}\fi

\bibitem{chaosbook}
P.~Cvitanovi\'{c}, R.~Artuso, R.~Mainieri, G.~Tanner, G.~Vattay, Chaos:
  Classical and Quantum, Niels Bohr Institute, Copenhagen, 2008, {\tt
  {http://ChaosBook.org}{ChaosBook.org}}.

\bibitem{Lan10}
Y.~Lan, Cycle expansions: From maps to turbulence,
  Commun.~Nonlinear~Sci.~Numer.~Simulat. 15~(3) (2010) 502--526.

\bibitem{davi99}
R.~L. Davidchack, Y.-C. Lai, Efficient algorithm for detecting unstable
  periodic orbits in chaotic systems, Phys.~Rev.~E 60~(5) (1999) 6172--6175.

\bibitem{Koh07}
Y.~W. Koh, K.~Takatsuka, Finding periodic orbits of higher-dimensional flows by
  including tangential components of trajectory motion, Phys.~Rev.~E 76~(6)
  (2007) 066205.

\bibitem{Biha89}
O.~Biham, W.~Wenzel, Characterization of unstable periodic orbits in chaotic
  attractors and repellers, Phys.~Rev.~Lett. 63~(8) (1989) 819--822.

\bibitem{Schm97}
P.~Schmelcher, F.~K. Diakonos, Detecting unstable periodic orbits of chaotic
  dynamical systems, Phys.~Rev.~Lett. 78~(25) (1997) 4733--4736.

\bibitem{Diak98}
F.~K. Diakonos, P.~Schmelcher, O.~Biham, Systematic computation of the least
  unstable periodic orbits in chaotic attractors, Phys.~Rev.~Lett. 81~(20)
  (1998) 4349--4352.

\bibitem{davi01}
R.~L. Davidchack, Y.-C. Lai, A.~Klebanoff, E.~M. Bollt, Towards complete
  detection of unstable periodic orbits in chaotic systems, Phys.~Lett.~A 287
  (2001) 99–104.

\bibitem{Lan04}
Y.~Lan, P.~Cvitanovi\'{c}, Variational method for finding periodic orbits in a
  general flow, Phys.~Rev.~E 69~(1) (2004) 016217.

\bibitem{Java88}
J.~Javanainen, J.~H. Eberly, Q.~Su, Numerical simulations of multiphoton
  ionization and above-threshold electron spectra, Phys.~Rev.~A 38~(7) (1988)
  3430--3446.

\bibitem{Haan94}
S.~L. Haan, R.~Grobe, J.~H. Eberly, Numerical study of autoionizing states in
  completely correlated two-electron systems, Phys.~Rev.~A 50~(1) (1994)
  378--391.

\bibitem{Maug09_1}
F.~Mauger, C.~Chandre, T.~Uzer, Strong field double ionization~: The phase
  space perspective, Phys.~Rev.~Lett. 102~(17) (2009) 173002.

\bibitem{Cork93}
P.~B. Corkum, Plasma perspective on strong field multiphoton ionization,
  Phys.~Rev.~Lett. 71~(13) (1993) 1994--1997.

\bibitem{Scha93}
K.~J. Schafer, B.~Yang, L.~F. DiMauro, K.~C. Kulander, Above threshold
  ionization beyond the high harmonic cutoff, Phys.~Rev.~Lett. 70~(11) (1993)
  1599--1602.

\bibitem{Beck08}
W.~Becker, H.~Rottke, Many-electron strong-field physics, Contemporary Physics
  49~(3) (2008) 199--223.

\bibitem{Kirk83}
S.~Kirkpatrick, C.~D. Gelatt, M.~P. Vecchi, Optimization by simulated
  annealing, Science 220~(4598) (1983) 671--680.

\bibitem{Metr53}
N.~Metropolis, A.~W. Rosenbluth, M.~N. Rosenbluth, A.~H. Teller, E.~Teller,
  Equation of state calculations by fast computing machines, J.~Chem.~Phys.
  21~(6) (1953) 1087--1092.

\bibitem{Maug09_2}
F.~Mauger, C.~Chandre, T.~Uzer, Strong field double ionization~: what is under
  the `knee'~?, J.~Phys.~B. 42 (2009) 165602.

\end{thebibliography}

\end{document}